\documentclass[11pt]{article}
\usepackage{fullpage,epsf,amssymb,amsfonts,amsmath,latexsym, graphicx,array,extarrows,mathrsfs,mathtools,amsthm,mathrsfs,subcaption}
\usepackage[normalem]{ulem}

\usepackage[usenames,dvipsnames,table]{xcolor}
\usepackage{multirow}
\usepackage{bm}
\usepackage{url}

\usepackage{authblk}
\usepackage{cite}

\usepackage{float}
\restylefloat{table}

\usepackage[utf8]{inputenc}

\numberwithin{equation}{section}

\title{\bf{The generalised infrared structure of the gluon propagator}}

\author[1]{Shirley Weishi Li\thanks{shirleyl@slac.stanford.edu}}
\author[2]{Peter Lowdon\thanks{peter.lowdon@polytechnique.edu}}
\author[3]{Orlando Oliveira\thanks{orlando@fis.uc.pt}}
\author[3]{Paulo J.~Silva\thanks{psilva@uc.pt}}

\affil[1]{\footnotesize{\textit{SLAC National Accelerator Laboratory, Stanford University, 2575 Sand Hill Rd, CA 94025, USA}}}
\affil[2]{\footnotesize{\textit{CPHT, CNRS, Ecole Polytechnique, Institut Polytechnique de Paris, Route de Saclay, 91128 Palaiseau, France}}}
\affil[3]{\footnotesize{\textit{CFisUC, Department of Physics, University of Coimbra, 3004-516 Coimbra, Portugal}}}

\date{}
\begin{document}

{\let\newpage\relax\maketitle}

\abstract
\noindent
Gauge theory correlators are potentially more singular in the infrared than those in non-gauge theories. We determine the implications that these singularities have on the spectrum of the theory, proving that the appearance of generalised poles implies the existence of on-shell states with fixed mass, but zero norm. For quantum chromodynamics these poles have direct relevance for the confinement of coloured states. Using lattice data for the Landau gauge gluon propagator we subsequently test for the presence of these poles, establishing that the data is indeed consistent with such a component.

\setcounter{page}{1}
\pagestyle{plain}

\newpage

\section{Introduction}  

\noindent
As with any quantum field theory (QFT), the fundamental field correlators in quantum chromodynamics (QCD) completely encode both the mathematical and physical properties of the theory. Analysing the structure of these objects is therefore essential for understanding these characteristics. Although the confinement of coloured states is a well-established feature of QCD, the precise mechanism which governs this phenomenon remains a deep open problem. Since the non-observability of colour occurs irrespective of whether the theory contains quarks or not, the structure of the gluon field correlators must therefore play a significant role in determining why confinement occurs in QCD, as opposed to other QFTs. Besides their purely theoretical relevance, gluon correlators are also of importance phenomenologically, entering into the calculation of many non-perturbative observables, including bound-state properties of hadrons such as decay widths and mass spectra~\cite{Alkofer:2000wg,Bender:2002as,Maris:2005tt,Roberts:2011cf,Haas:2013hpa}. \\

\noindent
For many years it has been understood that the gluon propagator, and in particular its low-momentum \textit{infrared} behaviour, is important for unravelling the non-perturbative structure of QCD~\cite{Mandula:1999nj}. Motivated by the idea that gluons could directly give rise to a linearly rising confinement potential, in the late 1970's Mandlestam~\cite{Mandelstam:1979xd} hypothesised that the non-perturbative gluon propagator may have a double massless pole as $p^{2} \rightarrow 0$. This idea was supported by phenomenological models at the time~\cite{Baker:1980gf,Baker:1980gd}, as well as initial calculations of the Dyson-Schwinger equations~\cite{Brown:1987ca}. Nevertheless, the earliest lattice data suggested that the infrared propagator was in fact describable by an ordinary massive pole, corresponding to some sort of effective excitation~\cite{Mandula:1987rh,Gupta:1987zc}. From a different perspective, motivated by the issues surrounding gauge copies, Gribov suggested that the gluon propagator could instead have a pair of complex conjugate poles~\cite{Gribov:1977wm}. Based on these ideas it was further argued by Zwanziger~\cite{Zwanziger:1989mf,Zwanziger:1991gz} that the propagator must also strictly vanish in the limit $p^{2} \rightarrow 0$, and up until a decade ago lattice evidence seemed to support this hypothesis. However, once it became feasible to perform calculations using large lattice volumes, results started to suggest that the propagator was actually non-vanishing at zero momentum~\cite{Cucchieri:2007md,Cucchieri:2007rg}. Despite further ideas being put forward to account for this behaviour~\cite{Dudal:2007cw,Dudal:2008rm}, including several recent studies that propose the existence of complex mass poles~\cite{Cyrol:2018xeq,Hayashi:2018giz,Binosi:2019ecz}, it still remains an open question as to how the propagator should behave in the infrared regime, and ultimately what theoretical implications this has for QCD. \\

\noindent
Although the precise non-perturbative structure of correlators in QCD remains largely unknown, progress can be made by first establishing the general properties of these objects. In local formulations of QFT quantum fields are defined to be operator-valued distributions, which among other things implies that any field correlator must be a distribution~\cite{Streater:1989vi,Haag:1992hx,Bogolyubov:1990kw}. It turns out that by combining this characteristic with the Poincar\'{e} transformation properties of the fields, and the physical assumption that the states in the theory have positive energy, this imposes significant analytic constraints on the structure of correlators, including the existence of a spectral representation~\cite{Bogolyubov:1990kw}. Despite the complications that can arise in gauge theories~\cite{Strocchi13}, analogous results also hold for the correlators in QCD, and in particular the gluon propagator~\cite{Lowdon:2017uqe, Lowdon:2018mbn}. The major goal of this work will be to fully elaborate the connection between the infrared structure of this representation, and the spectrum of the theory. Focussing on the specific case of Landau gauge, we will then use the Euclidean generalisation of this representation in order to test the feasibility of different infrared ans\"{a}tze using the high precision pure Yang-Mills lattice data from~\cite{Dudal:2018cli}, and ultimately help to shed new light on the non-perturbative gluon spectrum.

\section{The spectral structure of the gluon propagator}

\noindent
Irrespective of the matter content of the theory, it turns out that the gluon propagator has the following spectral representation~\cite{Lowdon:2017uqe, Lowdon:2018mbn}:
\begin{align}
\widehat{D}_{\mu\nu}^{ab\, F}(p) &=   i\int_{0}^{\infty} \frac{ds}{2\pi} \, \frac{\left[ g_{\mu\nu}\rho_{1}^{ab}(s) + p_{\mu}p_{\nu}\rho_{2}^{ab}(s) \right]}{p^{2}-s +i\epsilon} +\sum_{n=0}^{N} \left[ c_{n}^{ab} \, g_{\mu\nu} (\partial^{2})^{n} + d_{n}^{ab} \partial_{\mu}\partial_{\nu}(\partial^{2})^{n-1}\right]\delta^{4}(p),
\label{general_propagator_QCD_mom}
\end{align}
where $c_{n}^{ab}$ and $d_{n}^{ab}$ are complex coefficients, and $\partial_{\mu}= \tfrac{\partial}{\partial p^{\mu}}$. Defining: $D(p) = g^{\mu\nu}\widehat{D}_{\mu\nu}^{aa\, F}(p)$ and $\rho_{i} = \rho_{i}^{aa}$, it follows from Eq.~\eqref{general_propagator_QCD_mom} that the trace component of the Landau gauge gluon propagator $D(p)$ can be written
\begin{align}
D(p) =   3i\int_{0}^{\infty} \frac{ds}{2\pi}\frac{\rho_{1}(s)}{p^{2}-s+i\epsilon}   + \sum_{n=0}^{N+1}g_{n}(\partial^{2})^{n}\delta^{4}(p),
\label{trace}
\end{align}
with $g_{n}= 4c_{n}^{aa} + d_{n}^{aa}$. As detailed in~\cite{Lowdon:2017uqe} and references within, the representation in Eq.~\eqref{trace} follows from very broad assumptions, including the Poincar\'{e} covariance of the fields and the positivity of the energy-momentum spectrum. In particular, this implies that Eq.~\eqref{trace} holds independently of the non-perturbative subtleties associated with local quantisations such as Landau gauge. One can immediately recognise the first term in Eq.~\eqref{trace} as having the familiar-looking K\"{a}ll\'{e}n-Lehmann spectral form, whereas the second term is purely singular. The potential appearance of non-vanishing singular terms involving derivatives of $\delta^{4}(p)$ arises because in gauge theories such as QCD the gauge symmetry provides an obstacle to the locality of the theory~\cite{Strocchi13}. In order to consistently quantise the theory one therefore has two options: either allow non-local fields, or explicitly preserve locality. A general feature of local quantisations such as Landau gauge is that additional degrees of freedom are introduced into the theory, resulting in a space of states which includes unphysical negative-norm states such as ghosts~\cite{Nakanishi:1990qm}. It is precisely this loss of state-space positivity which prevents one from ruling out the existence of $\delta^{4}(p)$-derivative components~\cite{Bogolyubov:1990kw}.

\subsection{Generalised pole terms}   

\noindent
Just as the loss of state-space positivity allows for the possibility of purely singular terms, this feature also implies that the gluon spectral density $\rho_{1}(s)$ can potentially have more singular types of components. An important example of such components are \textit{generalised pole terms}~\cite{Oehme:1979bj,Lowdon:2015fig}:
\begin{align}
\delta^{(n)}(s-m_{n}^{2}) = \left(\tfrac{d}{ds}\right)^{n}\delta(s-m_{n}^{2}), \quad n \geq 1.
\label{gen_pole}
\end{align} 
Due to the appearance of derivatives in this expression these terms give rise to contributions in the propagator of the form $(p^{2}-m_{n}^{2} + i\epsilon)^{-n-1}$, which are more singular than ordinary massive poles. One can also demonstrate that each such term is associated with the existence of a finite-norm state $|\Psi_{n}\rangle$, satisfying the following conditions~\cite{Oehme:1979bj}:
\begin{align}
&(P^{2} - m_{n}^{2})^{n}|\Psi_{n}\rangle \neq 0, \label{off_shell} \\
&(P^{2} - m_{n}^{2})^{n+1}|\Psi_{n}\rangle = 0,  \label{on_shell}
\end{align}
where $P^{2}$ is the squared energy-momentum operator. Eq.~\eqref{off_shell} clearly emphasises that these states are off shell, and therefore represent unphysical degrees of freedom. The appearance of these types of states was first mentioned in~\cite{Oehme:1979bj}, although it was never established whether their presence is significant for the on-shell spectrum of the theory. In the remainder of this section we will address this question. \\

\noindent
Given $|\Psi_{n}\rangle$ satisfying Eqs.~\eqref{off_shell} and~\eqref{on_shell}, consider the state $|\widetilde{\Psi}_{n}\rangle$ defined by the repeated action ($n$ times) of the mass-shell operator $(P^{2}-m_{n}^{2})$:
\begin{align}
|\widetilde{\Psi}_{n}\rangle \equiv (P^{2} - m_{n}^{2})^{n}|\Psi_{n}\rangle. 
\end{align}
Due to Eq.~\eqref{on_shell} and the definition of $|\widetilde{\Psi}_{n}\rangle$ it follows that:
\begin{align}
(P^{2} - m_{n}^{2})|\widetilde{\Psi}_{n}\rangle = (P^{2} - m_{n}^{2})^{n+1}|\Psi_{n}\rangle =0,   
\end{align}
and therefore $|\widetilde{\Psi}_{n}\rangle$ defines an \textit{on-shell} state with mass $m_{n}$. Moreover, taking the inner product of $|\widetilde{\Psi}_{n}\rangle$ with itself one finds
\begin{align}
&\langle \widetilde{\Psi}_{n} |\widetilde{\Psi}_{n}\rangle =  \left(\langle \Psi_{n}|(P^{2} - m_{n}^{2})^{n}, \, (P^{2} - m_{n}^{2})^{n}|\Psi_{n}\rangle\right) \nonumber \\
& \ =  \langle \Psi_{n}|(P^{2} - m_{n}^{2})^{n-1}(P^{2} - m_{n}^{2})^{n+1}|\Psi_{n}\rangle =0,
\end{align}
where the last equality is implied by Eq.~\eqref{on_shell}. From this one can conclude that the appearance of an unphysical off-shell state $|\Psi_{n}\rangle$ necessarily implies the existence of an on-shell \textit{zero-norm} state $|\widetilde{\Psi}_{n}\rangle$, with mass $m_{n}$. Since $|\Psi_{n}\rangle$ are intrinsically connected to the appearance of generalised pole terms as in Eq.~\eqref{gen_pole}, this proves that despite the unconventional structure of these components, they nevertheless have a definite impact on the spectrum of the theory. For many years it has been understood that $\delta(s-m_{0}^{2})$ components in the spectral density with weights of differing signs correspond to on-shell states with either positive or negative norm, depending on the normalisation. Since $\delta^{(n)}(s-m_{n}^{2})$ with $n \geq 0$ are the only discrete mass terms that can appear in any spectral density, this analysis completes the classification of all possible on-shell states. Although the main focus of this work is the gluon propagator, these results apply generally to any QFT that has a space of states with an indefinite inner product.

\subsection{Asymptotic behaviour}  

\noindent
The relevance of the generalised pole terms in Eq.~\eqref{gen_pole} to the asymptotic behaviour of correlators was first recognised in~\cite{Strocchi:1975xz,Strocchi:1978ci}. In particular, it was proven that these terms can potentially cause the correlation strength between clusters of states to \textit{grow} with distance, a feature which does not occur in ordinary non-gauge theories~\cite{Streater:1989vi}. This characteristic is often referred to as a violation of the \textit{cluster decomposition property} (CDP). For clusters of coloured states this characteristic provides a mechanism which can guarantee their absence from the asymptotic spectrum, since a growth in correlation strength between coloured states would prevent the independent measurement of either of these states at large distances~\cite{Nakanishi:1990qm}. Establishing the spectral structure of the gluon propagator, or in fact the correlators involving any coloured fields, is therefore of particular importance for understanding confinement. In order to violate the CDP the spectral density must not only contain a component as in Eq.~\eqref{gen_pole}, but this component also needs to influence the behaviour of the correlator at $p^{2}=0$~\cite{Lowdon:2015fig,Strocchi:1975xz,Strocchi:1978ci}, which requires that $m_{n}=0$. The purely singular terms in Eq.~\eqref{general_propagator_QCD_mom} involving derivatives of $\delta^{4}(p)$ can also cause the CDP to be violated. However, in this case this occurs not because of the spectral characteristics, but due to the singular properties of the correlator itself.

\section{Lattice data fits}

\noindent
Although many different parametrisations of the gluon propagator have been considered over the years, the potential appearance of generalised poles has not yet been fully investigated. In the remainder of this section we will outline infrared fits to the lattice data of~\cite{Dudal:2018cli} testing for the existence of these terms. Since lattice calculations are inherently Euclidean, one must first consider the Euclidean generalisation of Eq.~\eqref{trace}. Under the assumption of the validity of the Euclidean analytic continuation, the Landau gauge gluon propagator has the form  
\begin{align}
D(p) =   3\int_{0}^{\infty} \frac{ds}{2\pi}\frac{\rho_{1}(s)}{p^{2}+s}   + \sum_{n=0}^{N+1}g_{n}(-\nabla^{2})^{n}\delta^{4}(p).
\label{trace_euclid}
\end{align}
Although the pole structure of the Euclidean and Minkowski propagators differ to one another, the spectral density itself remains unchanged. Nevertheless, since lattice simulations inherently possess a finite momentum resolution, the singular terms at $p=0$, if they exist, are presumably not detectable. We therefore disregarded these terms for the purpose of this analysis. \\

\noindent
In this analysis we used the $\beta = 6.0$ data of~\cite{Dudal:2018cli} with both $64^{4}$ and $80^{4}$ lattices, corresponding to a lattice spacing of $a=0.1016(25)\, \text{fm}$ and physical volumes of $(6.57 \, \text{fm})^{4}$ and $(8.21 \, \text{fm})^{4}$, respectively. Specific details of the sampling, gauge fixing and other definitions regarding the data can be found in~\cite{Dudal:2018cli} and references within. The first goal of our analysis was to test whether the following single-pole infrared propagator ans\"{a}tze could fit the lattice data up to some scale $p_{\text{max}}$
\begin{align}
D_{i}(p) = \frac{Z_{i}}{(p^{2} + m_{i}^{2})^{i+1}}, \quad i=0,1,2.
\label{ansatze}
\end{align}
The motivation for performing these specific fits is that it provides a way of testing whether the data is consistent with the appearance of different types of simple isolated poles in the gluon spectral density $\rho_{1}(s)$. If there does indeed exist such a pole at $s=m_{i}^{2}$, together with a gap $(m_{i}^{2},s_{c})$ in the spectrum up to some continuum threshold $s_{c}$, the infrared structure of the propagator will be dominated by these poles\footnote{A simple argument for this characteristic can be made by performing a small momentum expansion of the spectral representation, and observing that the continuum contributions are increasingly suppressed as the gap $(m_{i}^{2},s_{c})$ becomes larger.}. In particular, $D_{i}(p, m_{i})$ would provide a good fit to the data up to some $p_{\text{max}}< \sqrt{s_{c}}$. For these fits to make physical sense one requires that $p_{\text{max}}> m_{i}$. It is important to emphasise here that the ans\"{a}tze in Eq.~\eqref{ansatze} correspond to \textit{real} mass poles of differing orders, in contrast to the complex mass pole ans\"{a}tze often considered in the literature~\cite{Cyrol:2018xeq,Hayashi:2018giz,Binosi:2019ecz}. \\

\noindent
Due to the difficulty in precisely accessing the systematic uncertainties of the lattice data we decided to perform the fits using three different choices of uncertainties, with increasing levels of conservatism. For the first case we considered the statistical errors only, which are $\mathcal{O}(10^{-3})$. For the second case, we took into account both the statistical errors as well as an energy-dependent systematic error. This shape uncertainty was derived using gluon propagator data with different lattice spacings in~\cite{Duarte:2016iko}, and used an empirical function that can describe the differences between these results and hence capture the general shape changes which occur due to systematic effects. This approach is particularly well-suited to cases where the data is heavily influenced by spectrum shape effects, such as in the assignment of nuclear reactor flux errors for neutrino experiments~\cite{Capozzi:2015bpa}. Lastly, we adopted a data-independent choice for the systematic errors: we assumed that the propagator data could be modified by a fourth-order polynomial in $p^{2}$, with deviations at the $2.5$\% level. This was the most conservative choice for the systematic error since it allows for significant shape changes. In all of these fitting scenarios the goodness of fit was assessed using a Chi-squared minimisation procedure. The Chi-squared statistic definitions together with the fit results for the $64^{4}$ and $80^{4}$ lattice data are provided in the appendix. Since the $p=0$ data points in both the $64^{4}$ and $80^{4}$ data sets have a considerable systematic uncertainty relative to the other points, due to their calculation, we decided to omit $p=0$ from each of the final fits. Nevertheless, the omission of this point did not result in a significant modification to the parameter values in the fits.   \\

\begin{figure}
\centering
\includegraphics[width=0.5\columnwidth]{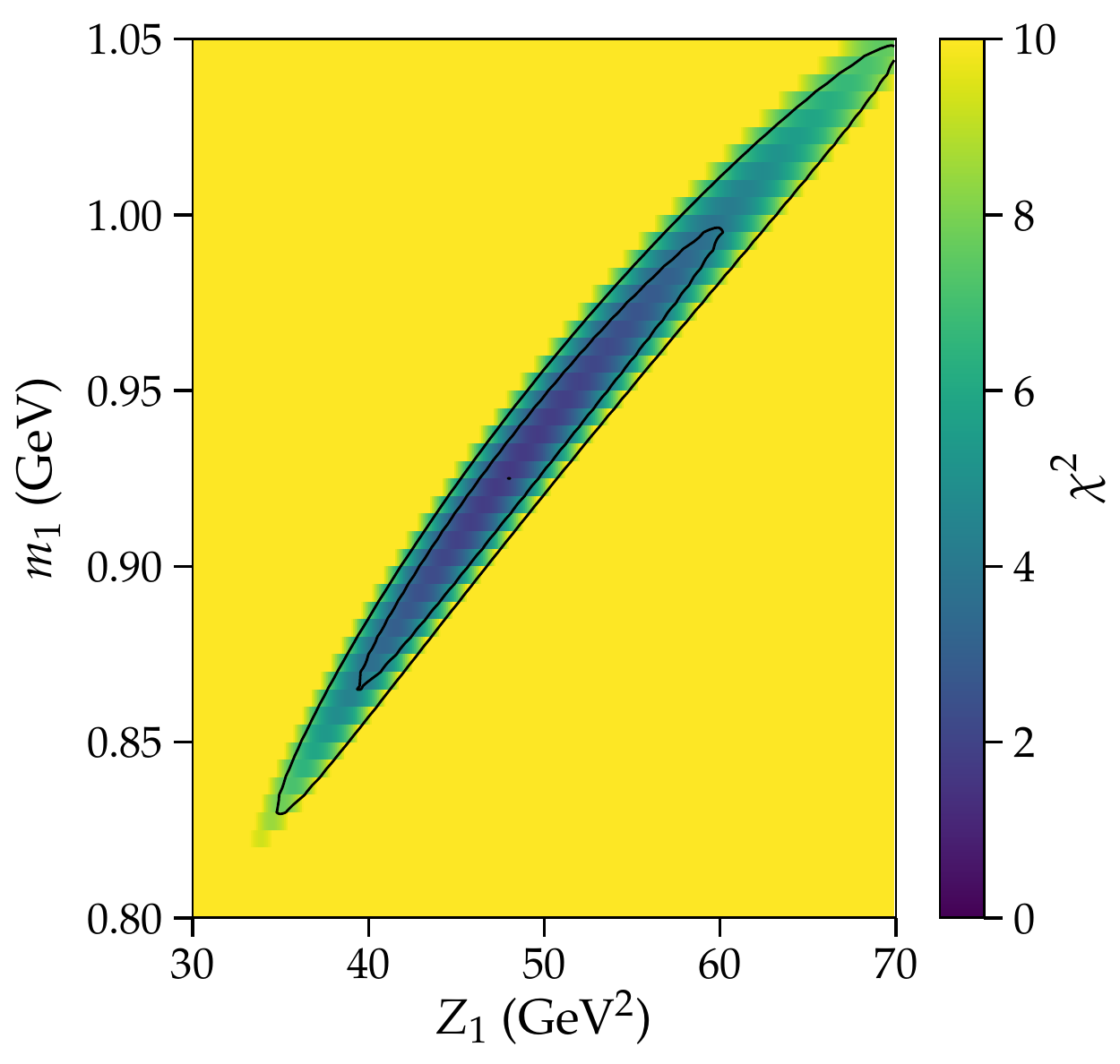}
\caption{Chi-squared map for the $80^4$ data $D_{1}(p)$ fit with statistical and polynomial shape uncertainty. The inner and outer black lines indicate the 1$\sigma$ and 2$\sigma$ uncertainty contours, respectively, and $p_{\text{max}}=  1\, \text{GeV}$.}
\label{fig:chi_square}
\end{figure}

\noindent
Comparing all of the fits in the different systematic error scenarios we found that $D_{1}(p)$ was the only ansatz which could provide a consistent fit to the data. In order to assess the robustness of these fits we performed a two-dimensional Chi-squared minimisation of the ans\"{a}tze parameters, and also tested the sensitivity of these fits to the momentum cutoff $p_{\text{max}}$. We found that neither $D_{0}(p)$ nor $D_{2}(p)$ provided a reliably stable fit for both data sets as $p_{\text{max}}$ varied, and that the fitted parameter values were highly sensitive to the choice of systematic errors. The $D_{1}(p)$ fits on the other hand resulted in a good fit to the data for both choices of (non-vanishing) systematic errors. Although the overall quality of the fits improved when using the larger volume data, the parameter values remained stable across the $64^{4}$ and $80^{4}$ fits. The Chi-squared values in $(Z_{1},m_{1})$ parameter space for the $80^4$ data with the polynomial shape choice of systematics are plotted in Fig.~\ref{fig:chi_square}. The parameter values at the global minimum of this fit are
\begin{align}
Z_{1} = 48  \substack{+12 \\ -8}  \ \text{GeV}^{2}, \quad m_{1} = 0.93  \substack{+0.07 \\ -0.06}    \ \text{GeV},
\label{result1}  
\end{align}
where the uncertainties indicate a 1$\sigma$ variation. We found that this fit remained convergent and physically consistent ($p_{\text{max}}> m_{1}$) up to $p_{\text{max}}=  1.5\, \text{GeV}$.  \\

\begin{figure}
\centering
\includegraphics[width=0.5\columnwidth]{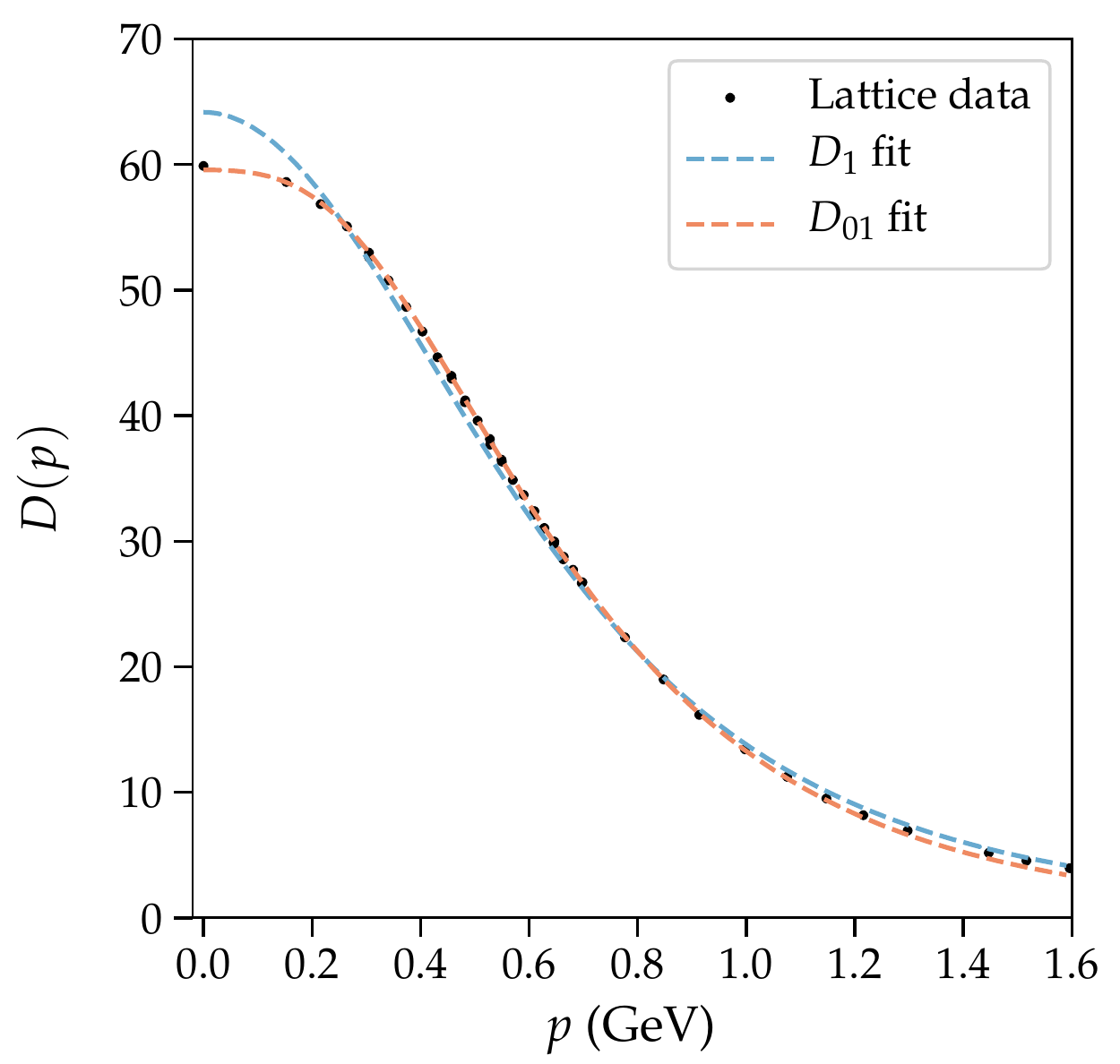}
\caption{Best fit plots of the $D_{1}(p)$ and $D_{01}(p)$ ans\"{a}tze together with the $80^4$ lattice data points. The statistical errors on the lattice data are very small and therefore not observable in the plot. The $p=0$ data point is plotted for reference, although it was not included in either of the fits due to its significant systematic uncertainty relative to the other data points.}
\label{fig:ansatze}
\end{figure}

\noindent
In principle it is also possible that the gluon spectral density could contain multiple types of poles. To investigate this possibility we therefore considered the simplest case of the two-term linear combinations of single pole components
\begin{align}
&D_{ij}(p) = \frac{Z_{i}}{(p^{2} + m_{i}^{2})^{i+1}} + \frac{Z_{j}}{(p^{2} + m_{j}^{2})^{j+1}}, 
\end{align}
where $i,j= 0,1,2$ and $i \neq j$. Overall we found that $D_{01}(p)$ provided a very good description of the lattice data, even when only statistical errors were included. The best fit was obtained with the $80^4$ data, yielding the corresponding generalised pole parameter values
\begin{align}
Z_{1} = 51  \substack{+8 \\ -6} \ \text{GeV}^{2}, \quad m_{1} = 0.88 \substack{+0.09 \\ -0.06} \ \text{GeV},
\label{result2}  
\end{align}
and: $Z_{0}=-3.4 \substack{+0.5 \\ -0.5}  \, \text{GeV}^{2}$, $m_{0}= 0.36 \substack{+0.04 \\ -0.04} \, \text{GeV}$. We found that this fit remained convergent and physically consistent ($p_{\text{max}}> \{m_{0}, m_{1}\}$) up to $p_{\text{max}}=  1.4 \, \text{GeV}$. Although the $D_{02}(p)$ and $D_{12}(p)$ ans\"{a}tze appeared to give a reasonable fit to the data based on the $\chi^2$/d.o.f. values alone, the fitted mass parameters were found to be degenerate, possessing a sensitivity to the momentum cutoff $p_{\text{max}}$. An important conclusion from this analysis is that the best fit $Z_{1}$ and $m_{1}$ parameter values from the $D_{01}(p)$ fit coincide (within errors) with those obtained in the $D_{1}(p)$ only fit. Moreover, adding a $D_{1}(p)$ component to either $D_{0}(p)$ or $D_{2}(p)$ ended up leading to a significant improvement in the quality of these fits. Taken together, these results support the hypothesis that the gluon propagator data is consistent with the existence of an infrared $D_{1}(p)$ contribution. The $D_{1}(p)$ and $D_{01}(p)$ ans\"{a}tze with the best fit parameter values in Eqs.~\eqref{result1} and~\eqref{result2}, respectively, are plotted together with the $80^4$ lattice data points in Fig.~\ref{fig:ansatze}.  \\ 

\noindent
Due to the spectral structure in Eq.~\eqref{trace_euclid}, a $D_{1}(p)$ contribution means that $\rho_{1}(s)$ contains a generalised component $\mathcal{Z}_{1}\delta'(s-m_{1}^{2})$, where $\mathcal{Z}_{1} = \tfrac{2}{3}\pi Z_{1}$. With the parameters given in Eq.~\eqref{result2} it follows from the previous theoretical analysis that the appearance of such a component implies the existence of an on-shell zero-norm state in the spectrum, with mass $m_{1} = 0.88 \, \text{GeV}$ and weight $\mathcal{Z}_{1}$. An important characteristic of this component is that it directly satisfies the Landau gauge superconvergence constraint~\cite{Lowdon:2018mbn,Oehme:1979ai,Cornwall:2013zra}: $\int ds \, \rho_{1}(s)= 0$, not due to any continuous oscillatory-type behaviour~\cite{Strauss:2012dg}, but because of the generalised singular properties of the $\delta'$ distribution. Since $m_{1}>0$ this also suggests that the gluon spectral density does not induce a singular enough behaviour near $p^{2}=0$ to result in a violation of the CDP. It could well be the case that a violation is indeed caused by the non-vanishing of one of the coefficients $g_{n}$ in Eq.~\eqref{trace_euclid}, but this remains to be seen. A non-violation of the CDP would imply that clusters of states created from single gluonic fields should decorrelate as they are separated further apart. Whilst this might initially appear surprising, one should note that the gluon correlator itself is not gauge-invariant, and so a non-violation of the CDP in this specific case does not contradict the physical expectation that asymptotic coloured states are prohibited.

\section{Conclusions}

\noindent
In locally quantised gauge theories the field correlators can potentially contain more singular generalised-pole contributions in their spectral densities. In this work we establish for the first time the connection between these types of components and the spectrum of the theory. In particular, we prove that the presence of generalised poles implies the existence of on-shell states in the spectrum with fixed mass, but zero norm. Despite their non-physicality, it turns out that these states can fundamentally alter the large-distance behaviour of correlators, which is important in the context of confinement. With these theoretical motivations in mind we analysed the high-precision Landau gauge gluon propagator lattice data from~\cite{Dudal:2018cli} in order to assess whether the data can accommodate generalised-pole contributions. Overall, we found that the data was consistent with the appearance of a massive single-derivative of delta component in the spectral density, and hence the existence of a corresponding zero-norm state in the spectrum. The non-vanishing mass of this state suggests that the spectral structure of the propagator alone is not sufficiently singular to guarantee the confinement of states involving single gluonic fields. Since generalised spectral components are also potentially present in other correlators, this work opens up a new direction for understanding the infrared structure of QCD.

\section*{Acknowledgements}
\noindent
S. Li was supported by the U.S. Department of Energy under contract DE--AC02--76SF00515, and P. Lowdon by the Agence Nationale de la Recherche under project No. ANR--18--ERC1--0002. P. J. Silva acknowledges support from FCT under the contracts SFRH/BPD/40998/2007 and SFRH/BPD/109971/2015, and both O. Oliveira and P. J. Silva acknowledge support from FCT under the contracts UID/FIS/04564/2016 and CERN/FIS-COM/0029/2017. This work was granted access to the HPC resources of the PDC Center for High Performance Computing at the KTH Royal Institute of Technology, Sweden, made available within the Distributed European Computing Initiative by the PRACE-2IP, receiving funding from the European Community's Seventh Framework Programme (FP7/2007-2013) under grand agreement no. RI-283493. The use of Lindgren has been provided under DECI-9 project COIMBRALATT. We acknowledge that the results of this research have been achieved using the PRACE-3IP project (FP7 RI312763) resource Sisu based in Finland at CSC. The use of Sisu has been provided under DECI-12 project COIMBRALATT2. We also acknowledge the Laboratory for Advanced Computing at the University of Coimbra (\url{http://www.uc.pt/lca}) for providing access to the HPC resource Navigator.

\newpage

\appendix 
\section{Goodness of fit results}
\label{appendix_a}

\noindent
The goodness of fit results for the $D_{i}(p)$ and $D_{ij}(p)$ ans\"{a}tze with the $64^{4}$ and $80^{4}$ lattice data of~\cite{Dudal:2018cli} are found in Tables~\ref{tab:chi_square_64} and~\ref{tab:chi_square_80}, respectively. In each fit a value of $p_{\text{max}} =1 \, \text{GeV}$ was initially chosen to determine whether a convergent and physically consistent ($p_{\text{max}}> m_{i}$) fit could be achieved. If not, the value of $p_{\text{max}}$ was lowered or raised until these conditions were satisfied. The values of $p_{\text{max}}$ in the tables reflect this final choice of cutoff. The $D_{02}(p)$ fits to both data sets were found to not be consistent when only statistical errors were included, regardless of the value of $p_{\text{max}}$. 
\ \\
\begin{table}[!ht]
\small
    \begin{center}
	\begin{tabular}{|c |c| c| c |c|}
	\hline
	 & \textbf{Stat. only} & \textbf{Stat. + Shape} & \textbf{Stat. + Poly.} \\
	 & $\chi_{1}^{2}$/d.o.f.  ($p_{\text{max}}$) & $\chi_{2}^{2}$/d.o.f.  ($p_{\text{max}}$) & $\chi_{3}^{2}$/d.o.f. ($p_{\text{max}}$) \\
	\hline
	$D_0(p)$ & $>150$ \ (0.7) & $>30$ (0.6) & 3.5 (0.7) \\  
	\hline
	$D_1(p)$ & $>90$ \ (1.0) & 3.5 \ (0.9) & 2.7 \ (1.0)\\
	\hline
	$D_2(p)$ & $>45$ \ (1.3) & $>45$ \ (1.3) & 3.1 \ (1.3) \\
	\hline 
	$D_{01}(p)$ & 1.9 \ (1.0) & 2.2 \ (1.0) & 2.1 \ (1.0) \\  
	\hline
	$D_{02}(p)$ & ---  & 2.7 \ (1.2) & 2.8 \ (1.3) \\
	\hline
	$D_{12}(p)$ & $>15$ \ (1.0) & 3.4 \ (1.0) & 2.5 \ (1.0) \\
	\hline    
	\end{tabular}
    \caption{Chi-squared fit results for $D_{i}(p)$ and $D_{ij}(p)$ under the different systematic error scenarios with the $64^4$ lattice data.}
    \label{tab:chi_square_64}
    \end{center}    
\end{table}

\begin{table}[!ht]
\small
    \begin{center}
	\begin{tabular}{|c |c| c| c |c|}
	\hline
	 & \textbf{Stat. only} & \textbf{Stat. + Shape} & \textbf{Stat. + Poly.} \\
	 & $\chi_{1}^{2}$/d.o.f.  ($p_{\text{max}}$) & $\chi_{2}^{2}$/d.o.f.  ($p_{\text{max}}$) & $\chi_{3}^{2}$/d.o.f. ($p_{\text{max}}$) \\
	\hline
	$D_0(p)$ & $>80$ \ (0.7)  & 7.3 \ (0.6) & 2.4 \ (0.7) \\  
	\hline
	$D_1(p)$ & $>25$ \ (1.0) & 1.8 \ (0.9) & 1.7 \ (1.0)\\
	\hline
	$D_2(p)$ & $>15$ \ (1.3) & $>15$  \ (1.3) & 1.8  \ (1.3) \\
	\hline 
	$D_{01}(p)$ & 1.1 \ (1.0) & 1.2 \ (1.0) & 1.2 \ (1.0) \\  
	\hline
	$D_{02}(p)$ & ---  & 1.6 \ (1.2)  & 1.6 \ (1.3) \\
	\hline
	$D_{12}(p)$ & 6.3 \ (1.2) & 1.7 \ (1.2)  & 1.5 \ (1.2)  \\
	\hline    
	\end{tabular}
    \caption{Chi-squared fit results for $D_{i}(p)$ and $D_{ij}(p)$ under the different systematic error scenarios with the $80^4$ lattice data.}
    \label{tab:chi_square_80}
    \end{center}    
\end{table}

\section{Chi-squared definitions}
\label{sec:chi-square}

\noindent
For each propagator ansatz tested we used three different choices for the systematic error, each of which requires a separate Chi-squared statistic definition in order to determine the goodness of fit. In the least conservative scenario for which we only consider the statistical errors, we used the following Chi-squared definition:
\begin{align}
\chi_{1}^{2} = (\bm{G}-\bm{D})^{\text{T}} \, \bm{C}^{-1} (\bm{G}-\bm{D}),
\label{chi1}
\end{align}
where $\bm{G}$ is the $n$-dimensional vector of lattice data at the discrete momenta $\{p_{1},..., p_{n}\}$, and $\bm{D}$ is the vector of the fitted propagator functional form evaluated at these same momenta. The covariance matrix $\bm{C}$ quantifies the statistical correlations between different field configurations in the data, and is highly diagonal. In the limit of zero cross-correlations $\bm{C}$ becomes the matrix of statistical variances at each lattice point, and Eq.~\eqref{chi1} reduces to the standard Chi-squared form. The minimum $\chi_{1}^2$ follows a $\chi^2$-distribution with $n-m$ degrees of freedom, where $m$ is the number of free parameters in the propagator ansatz. \\

\noindent
In the second scenario we took into account both the statistical error and an energy-dependent systematic error. The corresponding Chi-squared statistic is defined by
\begin{align}
\chi_{2}^{2} = (\bm{G}\cdot \bm{f}-\bm{D})^{\text{T}} \, \bm{C}^{-1} (\bm{G}\cdot \bm{f}-\bm{D}) + \frac{(a-\alpha)^2}{\sigma_{a}^2} + \frac{(b-\beta)^2}{\sigma_{b}^2},
\label{chi2}
\end{align}
where now the lattice data is modified by an effective shape function $\bm{f} = \bm{f}(\{p_{i}\},a,b)$. This fitting approach is particulary well-suited to cases where the data is heavily influenced by energy-dependent effects, for example in assigning a shape error to a nuclear reactor flux for neutrino experiments~\cite{Capozzi:2015bpa}. Since the purpose of $\bm{f}$ is to simulate the sort of shape changes that can occur due to systematic effects, such as the appearance of Gribov copies~\cite{Dudal:2018cli}, we used the propagator data from~\cite{Duarte:2016iko} in order to establish a reasonable functional form. In particular, $\bm{f}$ was established such that the shape of the data sets for the different lattice spacings in~\cite{Duarte:2016iko} could be transformed into one another after multiplication by this function. With this procedure the components of $\bm{f}$ were determined to have the functional form
\begin{align}
f_{i}(p_{i},a,b) = \frac{a+p_{i}^{2}}{ab+p_{i}^{2}},
\end{align}
and the priors: $\left(\alpha = 0.21,\sigma_{a}=0.016\right)$ and: $\left(\beta=1.0,\sigma_{b}=0.1\right)$ were found to cover the full range of shape differences observed in~\cite{Duarte:2016iko}. Due to the definition in Eq.~\eqref{chi2}, the minimum $\chi_{2}^2$ has $n-(m+2)$ degrees of freedom. In the most conservative systematic error scenario we used the following Chi-squared statistic      
\begin{align}
\chi_{3}^{2} = (\bm{G}\cdot \bm{g}-\bm{D})^{\text{T}} \, \bm{C}^{-1} (\bm{G}\cdot \bm{g}-\bm{D}) + \frac{|\bm{g}-\bm{\gamma}|^2}{\sigma_{g}^2},
\end{align}
where the systematic shape function $\bm{g}$ was chosen to be a forth-order polynomial in $p^{2}$ (with 4 free coefficients), with priors $\bm{\gamma}=1$ and $\sigma_{g}=0.025$ simulating a $2.5$\% variation from the data. The corresponding minimum $\chi_{3}^{2}$ has $n-(m+4)$ degrees of freedom.

\bibliographystyle{JHEP}
\bibliography{ref}

\end{document}